# Stückelberg path to pure de Sitter supergravity

Sukṛti Bansal,[1,*] Silvia Nagy,[2,†] Antonio Padilla,[3,4,‡] and Ivonne Zavala[5,§]

[1]*Institute for Theoretical Physics, TU Wien, Wiedner Hauptstraße 8-10/136, A-1040 Vienna, Austria*
[2]*Department of Mathematical Sciences, Durham University, Durham, DH1 3LE, United Kingdom*
[3]*School of Physics and Astronomy, University of Nottingham, Nottingham NG7 2RD, United Kingdom*
[4]*Nottingham Centre of Gravity, University of Nottingham, Nottingham NG7 2RD, United Kingdom*
[5]*Department of Physics, Swansea University, Swansea, SA2 8PP, United Kingdom*

We advance the study of pure de Sitter supergravity by introducing a finite formulation of unimodular supergravity via the super-Stückelberg mechanism. Building on previous works, we construct a complete four-dimensional action of spontaneously broken $\mathcal{N}=1$ supergravity to all orders in the Stückelberg fields, which allows for de Sitter solutions. The introduction of finite supergravity transformations extends the super-Stückelberg procedure beyond the second order, offering a recursive solution to all orders in the Goldstino sector. This work bridges the earlier perturbative approaches and the complete finite theory, opening new possibilities for de Sitter vacua in supergravity models and eventually string theory.

## I. INTRODUCTION

One of the most pressing challenges in fundamental physics today is the discrepancy between theoretical predictions and observations concerning the vacuum energy's size, known as the "cosmological constant problem" [1–4]. In the standard model of cosmology—the ΛCDM model (Lambda cold dark matter)—the accelerated expansion of the Universe is attributed to a tiny but constant energy density, Λ. This implies that our Universe is asymptotically de Sitter (dS), with a small vacuum energy. Although recent observations [5,6] suggest the possibility of a "dynamical vacuum energy" rather than a constant one [7], the core question remains: why is the cosmological constant so small or zero, especially when theoretical models predict much larger values?

The discovery of the Universe's accelerated expansion in 1998 triggered intense theoretical efforts to construct dS vacua within string theory and supergravity. The seminal work of Kachru-Kalosh-Linde-Trivedi (KKLT) in 2003 [8] proposed how to obtain dS vacua via an "uplifting" procedure. This method involves uplifting a supersymmetric anti–de Sitter (AdS) vacuum to a dS one by introducing an anti-D3 brane. The mechanism behind this uplift was later reinterpreted in a manifestly supersymmetric formalism, specifically through the use of a Volkov-Akulov (VA) Goldstino theory [9,10], which implements nonlinearly realized global supersymmetry, coupled to a supergravity background. This suggested the existence of a scalar-independent de Sitter supergravity. This work led to the development of a four-dimensional supergravity theory with spontaneously broken supersymmetry that admits dS vacua, commonly referred to as de Sitter supergravity. The first approach to construct this theory was undertaken in [11–13], where the key idea was the use of nilpotent constrained superfields, which eliminate the scalar component of the chiral multiplet and enforce nonlinear supersymmetry. A complete local supergravity action incorporating constrained superfields and nonlinear supersymmetry was later formulated in a series of works [14–19], which allowed for the construction of pure $\mathcal{N}=1$ supergravity models admitting dS solutions. An alternative approach was proposed in [20,21], where a Goldstino brane—a 3-brane object in superspace carrying the VA Goldstino—was coupled to minimal $\mathcal{N}=1$ off-shell supergravity. This method was shown to lead to the same four-dimensional action as the constrained superfield approach in [15], up to second order in the Goldstino.

In a different direction, inspired by classical unimodular gravity [22–28], a novel approach was introduced in [29] by three of us, where unimodular gravity was extended to supergravity using a super-Stückelberg mechanism.[1,2] We demonstrated in [38] that this super-Stückelberg approach yields the same four-dimensional action as the constructions in [20] (and thus [15]), again to second order in the Goldstino. In these previous works [29,38], we constructed the four-dimensional action perturbatively up to the second order in the Stückelberg fields. In the present paper, we take a significant step forward by constructing the super-Stückelberg action to all orders, providing the complete unimodular supergravity action. This represents a substantial advance in the development of a complete and consistent framework for de Sitter vacua in supergravity.

---

[*]Contact author: sukruti.bansal@tuwien.ac.at
[†]Contact author: silvia.nagy@durham.ac.uk
[‡]Contact author: antonio.padilla@nottingham.ac.uk
[§]Contact author: e.i.zavalacarrasco@swansea.ac.uk

This new approach to pure de Sitter supergravity has the potential to offer a new pathway to finding de Sitter vacua in perturbative string theory, being qualitatively different to traditional nilpotent superfield methods. For example, the Stückelberg fermion may be the Goldstino associated with broken superspace symmetries due to branes, as in [20], or else a proxy for higher form fields. Indeed, to the latter point, this work is an important step in generalizing Henneaux and Teitelboim's version of unimodular gravity [25] to the supersymmetric case. Such a theory would have potential application to the cosmological constant problem by inspiring a supersymmetric version of vacuum energy sequestering (VES) [39–45].

To introduce our construction strategy, we begin in Sec. II by discussing the case of unimodular gravity. Although our approach may appear unconventional compared to the standard treatment in the unimodular gravity literature, we present it as a gradual introduction to the more complex case of supergravity. Following this, in Sec. III, we move on to construct the full unimodular supergravity theory. In Sec. III A, we first introduce the finite supergravity transformations of a chiral superfield, which, to our knowledge, have not been presented previously. These transformations are essential for the subsequent section, where in Sec. IV, we develop the super-Stückelberg procedure to all orders. Finally, we conclude in Sec. V, while in Appendix A we give the combined infinitesimal diffeomorphism and local supersymmetry transformations of the supergravity fields. In Appendix B we summarize important identities and proofs used throughout the main text. Additionally, in Appendix C, we outline an alternative approach to our construction for further consideration.

## II. UNIMODULAR GRAVITY AND THE STÜCKELBERG PROCEDURE

As a warm-up, let us consider standard unimodular gravity, a restricted version of Einstein-Hilbert gravity for which the determinant of the metric is taken to be a constant [28,46–52].

---

[1]In recent years, the Stückelberg mechanism has been further extended to asymptotic symmetries [30–33].
[2]For other supersymmetric extensions of the unimodular theory, see [34–37].

Locally, the theory is equivalent to general relativity (GR), owing to the fact that the determinant of the metric can be set to any constant by a suitable choice of gauge in a neighborhood of any point in spacetime. The difference manifests itself globally, with the cosmological constant entering as an integration constant in unimodular gravity, rather than a fixed coupling constant, as in GR. The action for unimodular gravity can be written as follows:

$$S_{\text{UMG}} = \frac{1}{16\pi G_N} \int d^4x [\sqrt{-g}R - 2\Lambda(x)(\sqrt{-g} - \epsilon_0)], \quad (2.1)$$

where $\epsilon_0$ is a constant (often set to unity) and the Lagrange multiplier, $\Lambda(x)$, imposes the constraint on the determinant of the metric. This Lagrange multiplier transforms as a scalar under diffeomorphisms. This means that the last term in the action breaks diffeomorphisms explicitly. Indeed, the action is only invariant under transverse diffeomorphisms that preserve the determinant of the metric.

The full set of diffeomorphisms can be restored using a Stückelberg trick. This is done by introducing four new Stückelberg fields, $y^\mu(x)$, as if we were carrying out a passive coordinate transformation, $x^\mu \to y^\mu(x)$. The action becomes

$$S_{\text{UMG}} = \frac{1}{16\pi G_N} \int d^4x [\sqrt{-g}R - 2\Lambda(x)(\sqrt{-g} - \epsilon_0|\det J|)], \quad (2.2)$$

where the Jacobian $J^\mu{}_\nu = \partial y^\mu/\partial x^\nu$. The action is now manifestly invariant under diffeomorphisms $x^\mu \to x'^\mu(x)$, as long as the Stückelberg fields transform as scalars, $y^\mu(x) \to y'^\mu(x') = y^\mu(x)$. Variation of the action with respect to the Stückelberg fields now forces the Lagrange multiplier to be a constant, $\partial_\mu \Lambda = 0$. As such, it plays the role of a cosmological constant in the effective gravity equations. The system is equivalent to plain vanilla GR with a cosmological constant whose value is just an integration constant, presumably set by boundary conditions.

In order to align our discussion with what is to come for supergravity, let us redefine the Stückelberg fields using an exponential map,

$$y^\mu(x) = e^{\phi^\nu(x)\partial_\nu} x^\mu. \quad (2.3)$$

When there is an exponential coordinate transformation $y^\mu = e^{-K} x^\mu$, we have $\det J = 1 \cdot e^{-\bar{K}}$, with the operator $\bar{K} = K^\mu \bar{\partial}_\mu = (-1)^\mu \bar{\partial}_\mu K^\mu + (-1)^\mu \partial_\mu K^\mu$ acting to the left [cf. (B7)]. Setting $K^\mu = -\phi^\mu$, the $\det J$ appearing in action (2.2) is $= 1 \cdot e^{\bar{\phi}}$. After repeatedly integrating by parts, we can rewrite the action as

$$S_{\text{UMG}} = \frac{1}{16\pi G_N} \int d^4x [\sqrt{-g}(R - 2\Lambda) - \epsilon_0 e^{-\phi^\nu(x)\partial_\nu} \Lambda]. \quad (2.4)$$

This form of the action might have been anticipated by introducing the Stückelberg fields through an active, as opposed to a passive, transformation. This is because the scalar Lagrange multiplier transforms as

$$\Lambda \to e^{-\phi^\nu(x)\partial_\nu}\Lambda \qquad (2.5)$$

under an active coordinate transformation, where the transformation parameter is identified with the Stückelberg fields, $\phi^\mu(x)$. The action (2.4) is now readily obtained from the original unimodular action (2.2) by performing this active transformation on all fields.

The transformation law for the $\phi^\mu$ is nontrivial. Consider a general coordinate transformation, which, in the passive form, corresponds to $x^\mu \to x'^\mu = e^{\xi^\nu\partial_\nu}x^\mu$. If we express this as an active transformation on the various terms in the action (2.4), we see, for example, that $\Lambda(x) \to \Lambda'(x) = e^{-\xi^\nu(x)\partial_\nu}\Lambda(x)$ and $\phi^\mu(x) \to \phi'^\mu(x)$, where $\phi'^\mu$ is to be determined. Of course, the GR part of the action is automatically invariant under a general coordinate transformation. Focusing on the Stückelberg part of the Lagrangian, invariance requires that

$$e^{-\phi^\nu(x)\partial_\nu}\Lambda(x) \to e^{-\phi'^\nu(x)\partial_\nu}\Lambda'(x) = e^{-\phi^\nu(x)\partial_\nu}\Lambda(x). \qquad (2.6)$$

Using the fact that $\Lambda(x) = e^{\xi^\nu(x)\partial_\nu}\Lambda'(x)$, we immediately infer that

$$e^{-\phi'^\nu(x)\partial_\nu} = e^{-\phi^\nu(x)\partial_\nu}e^{\xi^\nu(x)\partial_\nu}. \qquad (2.7)$$

We now make use of the integral form of the Baker-Campbell-Hausdorff formula,[3] and work to linear order in $\xi$ to obtain

$$\phi'^\nu(x)\partial_\nu = \phi^\nu(x)\partial_\nu - \frac{\text{ad}_{[\phi^\nu(x)\partial_\nu]}}{1-e^{\text{ad}_{[\phi^\nu(x)\partial_\nu]}}}\xi^\nu(x)\partial_\nu$$

$$= \phi^\nu(x)\partial_\nu + \sum_{k=0}^\infty \frac{B_k^+(-1)^k\text{ad}^k_{[\phi^\nu(x)\partial_\nu]}}{k!}\xi^\nu(x)\partial_\nu, \qquad (2.8)$$

where $B_k^+$ are the Bernoulli numbers,

$$B_0^+ = 1, \quad B_1^+ = \frac{1}{2}, \quad B_2^+ = \frac{1}{6}, \quad B_3^+ = 0, \quad B_4^+ = -\frac{1}{30}, \ldots \qquad (2.9)$$

and $\text{ad}_X(Y) = [X, Y]$. It is relatively easy to show that $[\text{ad}_{[\phi^\nu(x)\partial_\nu]}]^k \xi^\nu(x)\partial_\nu = [L_\phi^k \xi^\nu]\partial_\nu$, where $L_\phi$ denotes the Lie derivative with respect to $\phi^\mu$. The (active) transformation law for $\phi^\mu$ can now readily be expressed as

$$\phi'^\mu(x) = \phi^\mu(x) - \frac{L_\phi}{1 - e^{L_\phi}}\xi^\mu(x). \qquad (2.10)$$

Although standard unimodular gravity is a useful framework in which to develop the tools we will use in the coming sections, its direct applications are limited. This is because it is equivalent to GR with a cosmological constant, at least at the classical level. The extension of these ideas to supergravity is more interesting, however, as it allows us to describe a supersymmetric action with de Sitter solutions that only break supersymmetry spontaneously. In the coming sections, we will see how this can be done consistently with a supergravity action, including Stückelberg fields at all orders.

## III. UNIMODULAR SUPERGRAVITY

We now turn our attention to supergravity and begin by reviewing the construction of unimodular supergravity via the super-Stückelberg procedure discussed in [29,38]. We start with the chiral superspace action for old minimal $\mathcal{N} = 1$ supergravity, working in the conventions of [55]

$$S = -\frac{6}{8\pi G_N}\int d^4x\, d^2\Theta\, \mathcal{E}\mathcal{R} + \text{H.c.}, \qquad (3.1)$$

where the components of the chiral supergravity superfield $\mathcal{R}$ are given by

$$\mathcal{R} = -\frac{1}{6}\Big\{M + \Theta(\sigma^\mu\bar{\sigma}^\nu\psi_{\mu\nu} - i\sigma^\mu\bar{\psi}_\mu + i\psi_\mu b^\mu)$$
$$+ \Theta^2\Big[\frac{1}{2}R + i\bar\psi^\mu\bar\sigma^\nu\psi_{\mu\nu} + \frac{2}{3}MM^* + \frac{1}{3}b_\mu b^\mu$$
$$- ie_a^\mu\mathcal{D}_\mu b^a + \frac{1}{2}\bar\psi\,\bar\psi\,M - \frac{1}{2}\psi_\mu\sigma^\mu\bar\psi_\nu b^\nu$$
$$+ \frac{1}{8}\varepsilon^{\mu\nu\rho\sigma}(\bar\psi_\mu\bar\sigma_\nu\psi_{\rho\sigma} + \psi_\mu\sigma_\nu\bar\psi_{\rho\sigma})\Big]\Big\}. \qquad (3.2)$$

Further, $\mathcal{E}$ is a chiral density superfield, generalizing the scalar density $\sqrt{-g}$ in GR whose components are

$$\mathcal{E} = \mathcal{F}_0 + \sqrt{2}\Theta\mathcal{F}_1 + \Theta\Theta\mathcal{F}_2, \quad \text{with}$$
$$\mathcal{F}_0 = \frac{1}{2}e,$$
$$\mathcal{F}_1 = \frac{i\sqrt{2}}{4}e\sigma^\mu\bar\psi_\mu,$$
$$\mathcal{F}_2 = -\frac{1}{2}eM^* - \frac{1}{8}e\bar\psi_\mu(\bar\sigma^\mu\sigma^\nu - \bar\sigma^\nu\sigma^\mu)\bar\psi_\nu, \qquad (3.3)$$

---

[3]Namely, given $e^Z = e^X e^Y$, the integral formula for $Z$ is given by $Z = X + (\int_0^1 \mathcal{B}(e^{\text{ad}_X}e^{t\text{ad}_Y})dt)Y$, where $\mathcal{B}(x) = \frac{x\log(x)}{x-1}$. Taking $X = \phi^\nu(x)\partial_\nu$ and $Y = \xi^\nu(x)\partial_\nu$ in (2.7) and working to linear order in $\xi^\mu(x)$, (2.7) leads to (2.8), upon using also that $\mathcal{B}(e^y) = \frac{y}{1-e^{-y}} = \sum_{k=0}^\infty \frac{B_k^+ y^k}{k!}$. See, e.g., [53] or [54] for details.

where the determinant of the vielbein, $e = \det e^{\mathbf{a}}_\mu$, $\psi_\mu(x)$ is the gravitino, and $b^\mu(x)$ and $M$ are the auxiliary fields in the old minimal supergravity model.

Under an infinitesimal and combined diffeomorphism and local supersymmetry transformation, which we will henceforth refer to as an infinitesimal supergravity transformation, $\mathcal{E}$ transforms as

$$\delta \mathcal{E} = -\partial_M[(-1)^M \Xi^M \mathcal{E}], \quad (3.4)$$

where

$$(-1)^M = \begin{cases} 1, & M = \mu, \\ -1, & M = \alpha, \end{cases} \quad (3.5)$$

and the superfield $\Xi^M$ is given by

$$\Xi^\mu = \xi^\mu + 2i\Theta\sigma^\mu\bar{\epsilon} + \Theta^2\bar{\psi}_\nu\bar{\sigma}^\mu\sigma^\nu\bar{\epsilon},$$

$$\Xi^\alpha = \epsilon^\alpha - i\Theta\sigma^\mu\bar{\epsilon}\psi^\alpha_\mu + \Theta^2\left[-i\omega_\mu^{\alpha\beta}(\sigma^\mu\bar{\epsilon})_\beta + \frac{1}{3}M^*\epsilon^\alpha - \frac{1}{2}\psi^\alpha_\nu(\bar{\psi}_\mu\bar{\sigma}^\nu\sigma^\mu\bar{\epsilon}) + \frac{1}{6}b_\mu(\epsilon\sigma^\mu\bar{\epsilon})^\alpha\right], \quad (3.6)$$

where $\epsilon$ is the parameter of local supersymmetry (SUSY) transformations and $\xi^\mu$ is the diffeomorphism parameter.

It is now convenient to introduce the following notation:

$$\Xi^M \equiv O^M{}_{\bar{N}}\boldsymbol{\xi}^{\bar{N}}, \quad (3.7)$$

where the index $\bar{N}$ runs over $(\mu, \alpha, \dot{\alpha})$, $\boldsymbol{\xi}^{\bar{N}}$ is a vector containing the diffeomorphism and supersymmetry transformation parameters

$$\boldsymbol{\xi}^{\bar{N}} \equiv \begin{pmatrix} \xi^\mu \\ \epsilon^\alpha \\ \bar{\epsilon}^{\dot{\alpha}} \end{pmatrix}, \quad (3.8)$$

and the components of the matrix $O^M{}_{\bar{N}}$ are given by

$$O^M{}_\nu = \begin{cases} \delta^\mu_\nu, & M = \mu \\ 0, & M = \alpha \end{cases}, \quad O^M{}_\beta = \begin{cases} 0, & M = \mu \\ \delta^\alpha_\beta + \frac{1}{3}\Theta^2 M^*\delta^\alpha_\beta, & M = \alpha \end{cases} \quad (3.9)$$

and

$$O^M{}_{\dot{\beta}} = \begin{cases} 2i\Theta^\beta\sigma^\mu_{\beta\dot{\beta}} + \Theta^2(\bar{\psi}_\nu\bar{\sigma}^\mu\sigma^\nu)_{\dot{\beta}}, & M = \mu \\ i\Theta^\beta\psi^\alpha_\mu\sigma^\mu_{\beta\dot{\beta}} + \Theta^2\left[-i\omega_\mu^{\alpha\beta}\sigma^\mu_{\beta\dot{\beta}} - \frac{1}{2}\psi^\alpha_\nu(\bar{\psi}_\mu\bar{\sigma}^\nu\sigma^\mu)_{\dot{\beta}} + \frac{1}{6}b_\mu\epsilon^{\alpha\gamma}\sigma^\mu_{\gamma\dot{\beta}}\right], & M = \alpha. \end{cases} \quad (3.10)$$

It is important to note that the $\Theta$ dependence of $\Xi^M$ is entirely contained in $O^M{}_{\bar{N}}$. For future use, let us also introduce the notation

$$\boldsymbol{\xi}^M = \Xi^M|_{\Theta=0} = O^M{}_{\bar{N}}|_{\Theta=0}\boldsymbol{\xi}^{\bar{N}} = \begin{pmatrix} \xi^\mu \\ \epsilon^\alpha \end{pmatrix}. \quad (3.11)$$

Following [29,38], we define the action for unimodular supergravity as

$$S = -\frac{6}{8\pi G_N}\int d^4x d^2\Theta\left[\mathcal{E}\mathcal{R} + \frac{1}{6}\Lambda(\mathcal{E} - \mathcal{E}_0)\right] + \text{H.c.}, \quad (3.12)$$

where we have introduced the Lagrange multiplier chiral superfield

$$\Lambda = \Lambda_0 + \sqrt{2}\Theta\Lambda_1 + \Lambda_2\Theta^2 \quad (3.13)$$

and the constant superfield[4]

$$\mathcal{E}_0 = \epsilon_0 + \frac{i}{2}m\Theta^2, \quad (3.14)$$

with $\epsilon_0$ and $m$ real constants. Action (3.12) has two chiral superfields $\mathcal{R}$ and $\Lambda$. We will show the supergravity transformation of a chiral superfield in Sec. III A. Varying action (3.12) with respect to $\Lambda$, we arrive at the constraint

$$\mathcal{E} = \mathcal{E}_0, \quad (3.15)$$

which is the SUSY analog of the unimodularity condition in GR. In components, (3.15) reads

---

[4]We take the spinor component of $\mathcal{E}_0$ to vanish for simplicity.

$$\frac{1}{2}e = \epsilon_0,$$

$$\frac{i\sqrt{2}}{4} e \sigma^\mu \bar{\psi}_\mu = 0,$$

$$-\frac{1}{2}eM^* - \frac{1}{8}e\bar{\psi}_\mu(\bar{\sigma}^\mu \sigma^\nu - \bar{\sigma}^\nu \sigma^\mu)\bar{\psi}_\nu = \frac{i}{2}m. \quad (3.16)$$

The action (3.12) is invariant under a restricted set of SUSY and diffeomorphism transformations, $\delta\mathcal{E} = 0$ [see Eq. (3.4)], exactly such that they preserve the conditions in (3.16).

### A. Finite supergravity transformation of a chiral superfield

We now introduce finite field-dependent supergravity transformations of a chiral superfield. A chiral superfield, such as $\Lambda$ defined in (3.13), transforms[5] under a combined infinitesimal active diffeomorphism and local supersymmetry transformation as

$$\Lambda(Z) \to \tilde{\Lambda}(Z) = \Lambda(Z) + \delta_\xi \Lambda(Z) = \Lambda(Z) - \Xi^M \partial_M \Lambda(Z), \quad (3.17)$$

with $Z^M = (x^\mu, \Theta^\alpha)$ as the chiral superspace coordinates, $\Xi^M$ defined in (3.6), and the relation between $\Xi^M$ and $\boldsymbol{\xi}$ given in (3.7). Note that $\Xi^M$ depends not only on the diffeomorphism and SUSY parameters $\xi^\mu$, $\epsilon^\alpha$, and $\bar{\epsilon}^{\dot\alpha}$, but also on the fields in the supergravity multiplet, which we now denote collectively by

$$\varphi_{sg} = (e_\mu^a, \psi_\mu^\alpha, b_\mu, M). \quad (3.18)$$

The infinitesimal passive transformation corresponding to (3.17), i.e.,

$$Z^M \to \tilde{Z}^M = Z^M - \delta_\xi Z^M = Z^M + \Xi^M, \quad (3.19)$$

is contractible.[6] Therefore, its finite version can be obtained via exponentiation,

$$Z^M \to Z'^M = e^{-\delta_\xi} Z^M. \quad (3.20)$$

On taking this finite passive transformation, identity (B6) gives us the finite active supergravity transformation of the chiral superfield $\Lambda(Z)$ as

$$\Lambda(Z) \to \Lambda'(Z) = e^{\delta_\xi} \Lambda(Z). \quad (3.21)$$

---

[5]The transformation of a supersmooth scalar function in superspace is analogous to the transformation of a scalar field in curved spacetime.

[6]A coordinate transformation is called "contractible" if it can be deformed continuously to the identity transformation (see [56]).

The composition of two finite transformations follows the Baker-Campbell-Hausdorff formula

$$e^{\delta_{\xi_1}} e^{\delta_{\xi_2}} = e^{\delta_{\xi_1} + \delta_{\xi_2} + \frac{1}{2}[\delta_{\xi_1}, \delta_{\xi_2}] + \frac{1}{12}[\delta_{\xi_1},[\delta_{\xi_1}, \delta_{\xi_2}]] - \frac{1}{12}[\delta_{\xi_2},[\delta_{\xi_1}, \delta_{\xi_2}]] + \cdots}, \quad (3.22)$$

where the ... denote higher-order commutators. The group structure of supergravity transformations implies that the commutator of two infinitesimal transformations should give another infinitesimal transformation, i.e.,

$$[\delta_{\xi_1}, \delta_{\xi_2}] = \delta_{\xi_3(\xi_1,\xi_2)} = -\Xi_3^M \partial_M. \quad (3.23)$$

Acting on a chiral superfield, we have

$$\begin{aligned}
\delta_{\xi_1}\delta_{\xi_2}\Lambda &= -\delta_{\xi_1}(\Xi_2^M \partial_M \Lambda) \\
&= -(\delta_{\xi_1}\Xi_2^M)\partial_M \Lambda - \Xi_2^M \partial_M \delta_{\xi_1}\Lambda \\
&= -\left(\int d^6 Y \frac{\delta \Xi_2^M[\varphi_{sg}(Z)]}{\delta \varphi_{sg}(Y)} \delta_{\xi_1}\varphi_{sg}(Y)\right)\partial_M \Lambda \\
&\quad + \Xi_2^N \partial_N(\Xi_1^M \partial_M \Lambda) \\
&= -\frac{\partial \Xi_2^M}{\partial \varphi_{sg}} \delta_{\xi_1}\varphi_{sg}\partial_M \Lambda + (\Xi_2^N \partial_N \Xi_1^M)\partial_M \Lambda \\
&\quad + \Xi_2^N \Xi_1^M \partial_M \partial_N \Lambda,
\end{aligned} \quad (3.24)$$

where in the third line the integral is over the chiral superspace, with $Y^M = (x^\mu, \Theta^\alpha)$. Crucially, unlike, for example, pure diffeomorphism transformations, we must take into account the variations of the supergravity fields appearing in the transformation rules. The expressions for $\delta_\xi \varphi_{sg}$ are given in Appendix A. On computing the commutator in Eq. (3.23) using result (3.24), we get

$$\begin{aligned}
\Xi_3^M &= \Xi_1^N \partial_N \Xi_2^M - \Xi_2^N \partial_N \Xi_1^M + \frac{\partial \Xi_2^M}{\partial \varphi_{sg}}\delta_{\xi_1}\varphi_{sg} - \frac{\partial \Xi_1^M}{\partial \varphi_{sg}}\delta_{\xi_2}\varphi_{sg} \\
&= [\Xi_1, \Xi_2]_{SL} + \delta_{\xi_1}\Xi_2^M - \delta_{\xi_2}\Xi_1^M,
\end{aligned} \quad (3.25)$$

where we introduced $[,]_{SL}$, which can be seen as the superspace generalization of the standard Lie bracket for vector fields. This is a natural generalization of the corresponding case in general relativity. The additional terms are a consequence of the fact that, unlike in general relativity, our transformation parameters now depend on supergravity fields. We remark that (3.25) is just the supergravity version of deformed brackets for field-dependent parameters, which have appeared in [30–32,57–60].

## IV. SUPER-STÜCKELBERG PROCEDURE TO ALL ORDERS

In [29,38], the super-Stückelberg trick was performed up to the second order in the SUSY and diffeomorphism transformation parameters—both via active and passive

transformations—by promoting them to Stückelberg fields, $\xi \to \phi$, namely,

$$\xi^{\bar{N}} = \begin{pmatrix} \xi^\mu \\ \epsilon^\alpha \\ \bar{\epsilon}^{\dot{\alpha}} \end{pmatrix} \quad \longrightarrow \quad \phi^{\bar{N}} = \begin{pmatrix} \phi^\mu \\ \zeta^\alpha \\ \bar{\zeta}^{\dot{\alpha}} \end{pmatrix}. \quad (4.1)$$

In analogy with (3.7) and (3.11), we can also define the objects

$$\Phi^M \equiv O^M{}_{\bar{N}} \phi^{\bar{N}} \quad (4.2)$$

and

$$\phi^M = \Phi^M|_{\Theta=0} = O^M{}_{\bar{N}}|_{\Theta=0} \phi^{\bar{N}} = \begin{pmatrix} \phi^\mu \\ \zeta^\alpha \end{pmatrix}. \quad (4.3)$$

It was then shown that the resulting action admits a maximally symmetric solution, with the cosmological constant given by

$$\Lambda_{\text{eff}} = \Lambda_2 - \frac{1}{3} m^2, \quad (4.4)$$

where $\Lambda_2$ is the boundary condition for the $\Lambda_2$ component of the Lagrange multiplier superfield $\Lambda$ introduced in Eq. (3.13). Similar to the constrained superfield literature [13,14,16–18,20,21], the value of $\Lambda_{\text{eff}}$ can have either sign. In particular, we can have $\Lambda_{\text{eff}} > 0$ corresponding to a de Sitter solution that spontaneously breaks the underlying supersymmetry. Furthermore, in [38], the action to cubic order was obtained and compared with that in [20], giving matching results.

In this section, we generalize the results of [29,38] to show how to perform the super-Stückelberg procedure described above to all orders, such that the solutions with $\Lambda_{\text{eff}}$ above can be obtained from an action that has the full diffeomorphism and local SUSY symmetries. We take an all-order active transformation, as given in Eq. (3.21). It can be seen that the last term of action (3.12) does not remain invariant under the above transformation. In order to restore the broken supergravity symmetries, we now perform the Stückelberg procedure by sending

$$e^{\delta_\xi} \Lambda(Z) \to e^{\delta_\phi} \Lambda(Z), \quad (4.5)$$

where the $\delta$ used on the lhs is the standard variation that acts on all fields, whereas the $\delta$ used on the rhs does not act on Stückelberg fields. The need for two different versions of the variation arises because of the Stückelberg procedure, as follows. We know that $e^{\delta_\xi} = \sum_{n=0}^\infty \frac{(\delta_\xi)^n}{n!}$. Before the introduction of the Stückelberg field $\phi$, there is no $\phi$ for $\delta_\xi$ to act on. After $\xi$ is promoted to $\phi$, a general variation $\delta$ is supposed to act on $\phi$ as well.[7] However, due to the Stückelberg procedure, $e^{\delta_\phi}$ in Eq. (4.5) is $e^{\delta_\phi} = \sum_{n=0}^\infty \frac{(\delta_\xi)^n}{n!}|_{\xi=\phi} \neq \sum_{n=0}^\infty \frac{(\delta_\phi)^n}{n!} = e^{\delta_\phi}$.

In the end, we arrive at the final form of the Stückelberg action,

$$S = -\frac{6}{8\pi G_N} \int d^6 Z \left( \mathcal{E} R + \frac{1}{6} \Lambda \mathcal{E} - \frac{1}{6} \mathcal{E}_0 e^{\delta_\phi} \Lambda \right) + \text{H.c.} \quad (4.6)$$

Under a suitable transformation law for the Stücklelberg fields, this action is invariant under full diffeomorphisms and local SUSY transformations. We now derive the form of that transformation law explicitly.

### A. Supergravity transformation of Stückelberg fields

We now require action (4.6) to be invariant under supergravity transformations, which will provide us with the transformation rules for the Stückelberg fields $\phi^\mu$, $\zeta^\alpha$, and $\bar{\zeta}^{\dot{\alpha}}$ to all orders in the fields. The first two terms of action (4.6) are already invariant by construction. Let us denote the final term by

$$S_\phi = \int d^6 Z \, \mathcal{L}_\phi(Z), \quad (4.7)$$

where

$$\mathcal{L}_\phi(Z) = \frac{1}{8\pi G_N} e^{\delta_\phi} \Lambda(Z) \mathcal{E}_0. \quad (4.8)$$

We now take the (passive) finite supercoordinate transformation

$$Z^M \to Z'^M = e^{-\delta_\xi} Z^M, \quad (4.9)$$

so that the action transforms as

$$S_\phi \to S'_\phi = \int d^6 Z' \, \mathcal{L}'_\phi(Z') = \int d^6 Z |\text{Ber} J| \mathcal{L}'_\phi(Z'), \quad (4.10)$$

where

$$\text{Ber} J = 1 \cdot e^{-\bar{\delta}_\xi}, \quad (4.11)$$

and we refer the reader to (B7) for more details. Performing integration by parts [cf. (B8)], we can rewrite $S'_\phi$ as

$$S'_\phi = \int d^6 Z \, e^{\delta_\xi} \mathcal{L}'_\phi(Z'). \quad (4.12)$$

---

[7]The Stückelberg procedure requires that $\delta_\phi$ acts on exactly the same fields that $\delta_\xi$ acts on before the introduction of the Stückelberg fields. Therefore, the $\delta$ shown on the rhs does not act on Stückelberg fields $\phi$.

Using the identity (B5), we find that

$$\mathcal{L}'_\phi(Z') = e^{-\delta_\xi}\mathcal{L}'_\phi(Z), \quad (4.13)$$

which gives $S'_\phi = \int d^6 Z \mathcal{L}'_\phi(Z)$. For the action to remain invariant, we therefore require that

$$\mathcal{L}'_\phi(Z) = \mathcal{L}_\phi(Z). \quad (4.14)$$

In view of (4.8), this translates to

$$e^{\delta'_\phi}\Lambda'(Z) = e^{\delta_\phi}\Lambda(Z)$$
$$\Rightarrow e^{\delta'_\phi}e^{\delta_\xi}\Lambda(Z) = e^{\delta_\phi}\Lambda(Z) = e^{\delta_\phi}e^{-\delta_\xi}e^{\delta_\xi}\Lambda(Z), \quad (4.15)$$

where we used the fact that $\Lambda$ transforms as a chiral superfield [see (3.21)]. Finally, we read off

$$e^{\delta'_\phi} = e^{\delta_\phi}e^{-\delta_\xi}, \quad (4.16)$$

which is strongly reminiscent of the corresponding formula given by Eq. (2.7) in the pure gravity case. Using the integral Baker-Campbell-Hausdorff formula (see footnote 3), we obtain the following transformation law for the Stückleberg fields:

$$\delta'_\phi = \delta_\phi - \left[\int_0^1 \mathcal{B}(e^{\mathrm{ad}_{\delta_\phi}}e^{-t\mathrm{ad}_{\delta_\xi}})dt\right]\delta_\xi, \quad (4.17)$$

with

$$\mathcal{B}(x) = \frac{x\log(x)}{x-1}. \quad (4.18)$$

To get an explicit form for the transformation law to leading order, we truncate (4.17) beyond the linear order in transformation parameters, which gives

$$\delta'_\phi = \delta_\phi - \mathcal{B}(e^{\mathrm{ad}_{\delta_\phi}})\delta_\xi. \quad (4.19)$$

Using the fact that

$$\mathcal{B}(e^y) = \frac{y}{1-e^{-y}} = \sum_{k=0}^\infty \frac{B_k^+ y^k}{k!}, \quad (4.20)$$

where $B_k^+$ are the Bernoulli numbers given in Eq. (2.9), we arrive at the following expression:

$$\delta'_\phi = \delta_\phi - \sum_{k=0}^\infty \frac{B_k^+}{k!}\mathrm{ad}^k_{\delta_\phi}\delta_\xi. \quad (4.21)$$

As a sanity check, we can compare this expression with Eq. (2.8) in the case of GR, where we would identify $\delta_\phi$ with $-\phi^\nu\partial_\nu$ when acting on a scalar field. Furthermore,

since

$$\mathrm{ad}^k_{\delta_\phi}\delta_\xi = [\underbrace{\delta_\phi,\ldots[\delta_\phi,[\delta_\phi,\delta_\xi]]\ldots}_{k\text{ times}}], \quad (4.22)$$

where $\delta_\phi$ appears $k$ times, the value of $k$ in a given term in Eq. (4.21) represents the order of the Stückelberg field $\phi$ in it, as will be seen explicitly in the next section.

In Eq. (4.21), $\delta'_\phi$ is obtained on demanding the invariance of our Stückelberg supergravity action (4.6) under general supercoordinate transformation. Now, we derive the expression for $\delta'_\phi$ explicitly by performing the supergravity transformation $\pmb{\delta}_\xi$ of $\delta_\phi$. For this, we first need the expression for $\delta_\phi$. Recall that when $\delta_\phi$ acts on a chiral superfield, it is written as [as can also be seen by performing the Stückelberg procedure on Eq. (3.17)]

$$\delta_\phi = -\Phi^M \partial_M. \quad (4.23)$$

Plugging the expression for $\Phi^M$ from Eq. (4.2) into the above equation, we get the expression for $\delta_\phi$,

$$\delta_\phi = -O^M{}_{\bar{N}}\phi^{\bar{N}}\partial_M. \quad (4.24)$$

Under the supergravity transformation $\pmb{\delta}_\xi$, $\delta_\phi$ becomes $\delta'_\phi$, as follows[8]:

$$\begin{aligned}\delta_\phi \to \delta'_\phi &= -(O')^M{}_{\bar{N}}(\phi')^{\bar{N}}\partial'_M \\ &= \delta_\phi + \pmb{\delta}_\xi(\delta_\phi) \\ &= \delta_\phi + \delta_\xi(\delta_\phi) - O^M{}_{\bar{N}}\delta_\xi\phi^{\bar{N}}\partial_M \\ &= \delta_\phi + [\delta_\xi,\delta_\phi] - O^M{}_{\bar{N}}\delta_\xi\phi^{\bar{N}}\partial_M \\ &= \delta_\phi - \mathrm{ad}_{\delta_\phi}\delta_\xi - O^M{}_{\bar{N}}\delta_\xi\phi^{\bar{N}}\partial_M, \end{aligned} \quad (4.25)$$

where in the second-last line we have used identity (B1). Comparing this result (4.25) with (4.21), we find that

$$\begin{aligned}O^M{}_{\bar{N}}\delta_\xi\phi^{\bar{N}}\partial_M &= -\mathrm{ad}_{\delta_\phi}\delta_\xi + \sum_{k=0}^\infty \frac{B_k^+}{k!}\mathrm{ad}^k_{\delta_\phi}\delta_\xi \\ &= -2B_1^+\mathrm{ad}_{\delta_\phi}\delta_\xi + \sum_{k=0}^\infty \frac{B_k^+}{k!}\mathrm{ad}^k_{\delta_\phi}\delta_\xi \\ &= \sum_{k=0}^\infty \frac{B_k^+}{k!}(1-2\delta_{k,1})\mathrm{ad}^k_{\delta_\phi}\delta_\xi \\ &= \sum_{k=0}^\infty \frac{B_k^-}{k!}\mathrm{ad}^k_{\delta_\phi}\delta_\xi, \end{aligned} \quad (4.26)$$

where $B_k^-$ differ from $B_k^+$ only for $k=1$: $B_k^- = -1/2$ and $B_k^+ = 1/2$. Using the well-known formula

---

[8]Recall the distinction between $\pmb{\delta}$ and $\delta$ introduced in (4.5).

$$\sum_{k=0}^{\infty} \frac{B_k^- y^k}{k!} = \frac{y}{e^y - 1}, \tag{4.27}$$

we also find a closed form for (4.26), given by

$$O^M{}_{\bar{N}} \delta_{\xi} \phi^{\bar{N}} \partial_M = \frac{\mathrm{ad}_{\delta_\phi}}{e^{\mathrm{ad}_{\delta_\phi}} - 1} \delta_{\xi}. \tag{4.28}$$

The crucial point is that the rhs of the above is a sum of nested commutators or transformations (where the Stückelberg field is not transformed, i.e., it acts like a parameter). Our goal is to find the explicit expressions for the supergravity transformations of the Stückelberg fields, viz. $\delta_{\xi}\phi^\mu, \delta_{\xi}\zeta^\alpha$, and $\delta_{\xi}\bar\zeta^{\dot\alpha}$. Simply taking the $\Theta = 0$ component of the lhs of (4.28) gives us these expressions, as explained below. Using Eqs. (3.9), (3.10), and (4.3), we find that

$$\begin{aligned}(O^M{}_{\bar{N}} \delta_{\xi} \phi^{\bar{N}} \partial_M)|_{\Theta=0} \\ &= (O^M{}_{\nu} \delta_{\xi}\phi^\nu \partial_M + O^M{}_\beta \delta_{\xi}\phi^\beta \partial_M + O^M{}_{\dot\beta} \delta_{\xi}\phi^{\dot\beta} \partial_M)|_{\Theta=0} \\ &= (\delta^\mu_\nu \delta_{\xi}\phi^\nu \partial_\mu + \delta^\alpha{}_\beta \delta_{\xi}\phi^\beta \partial_\alpha + 0)|_{\Theta=0} \\ &= \delta_{\xi}\phi^M \partial_M. \end{aligned} \tag{4.29}$$

The above relation represents the supergravity transformation of $\phi^M$, which consists of $\phi^\mu$ and $\zeta^\alpha$. Since $\delta_{\xi}\bar\zeta^{\dot\alpha}\partial_{\dot\alpha} = (\delta_{\xi}\zeta^\alpha \partial_\alpha)^\dagger$, from the above relation we can determine the transformation $\delta_{\xi}$ of all three Stückelberg fields—$\phi^\mu, \zeta^\alpha$, and $\bar\zeta^{\dot\alpha}$. On equating Eq. (4.29) with the $\Theta = 0$ component of the rhs of Eq. (4.26), we have the final expression for the infinitesimal supergravity transformation of the Stückelberg fields,

$$\delta_{\xi}\phi^M \partial_M = \sum_{k=0}^{\infty} \frac{B_k^-}{k!} \mathrm{ad}^k_{\delta_\phi} \delta_{\xi}|_{\Theta=0}. \tag{4.30}$$

In the rhs above, a $k$th-order term is a nested commutator of the $k$th order [cf. Eq. (4.22)], which can be a long and bulky expression to evaluate and simplify. We are interested in finding concise and explicit expressions for the $k$th-order terms. This can be done recursively. We do so in the next section.

### B. Recursive transformation of Stückelberg fields order-by-order

We can now use relation (4.30) to find the explicit expressions for the infinitesimal supergravity transformation of the Stückelberg fields at all orders in $\phi^{\bar{M}}$. Besides the series expression in Eq. (4.30), $\delta_{\xi}\phi^M \partial_M$ can be expressed order-by-order in $k$, as follows:

$$\delta_{\xi}\phi^M \partial_M = \delta_{\xi}^{(0)}\phi^M \partial_M + \delta_{\xi}^{(1)}\phi^M \partial_M + \cdots + \delta_{\xi}^{(k)}\phi^M \partial_M + \cdots, \tag{4.31}$$

where $\delta^{(k)}$ denotes the supergravity transformation of $\phi^M$ at the $k$th order in Stückelberg fields and linear in $\xi^{\bar{M}}$. It will be convenient to define a $k$th-order transformation parameter $\xi^{(k)}(\phi, \xi)$ via the following action on a chiral superfield: $\delta_{\xi^{(k)}(\phi,\xi)}{}^9 = -\Xi^{(k)M}\partial_M = \mathrm{ad}^k_{\delta_\phi}\delta_{\xi}$ [cf. (3.17)], where $\Xi^{(k)M} = O^M{}_{\bar{N}}\xi^{(k)\bar{N}}$ [cf. (3.7)]. From Eq. (4.30), it follows that $\delta^{(k)}_{\xi}\phi^M = -\frac{B_k^-}{k!}\Xi^{(k)M}|_{\Theta=0}$.

From the zeroth order of Eq. (4.30), i.e., for $k = 0$, we have

$$\begin{aligned}\delta^{(0)}_{\xi}\phi^M \partial_M &= \delta_{\xi}|_{\Theta=0} = \delta_{\xi^{(0)}(\phi,\xi)}|_{\Theta=0} = -\Xi^{(0)M}\partial_M|_{\Theta=0} \\ &= -O^M{}_{\bar{N}}\xi^{(0)\bar{N}}\partial_M|_{\Theta=0} = -\xi^{(0)M}\partial_M. \end{aligned} \tag{4.32}$$

Note that $\Xi^{(0)M} = \Xi^M$ and $\xi^{(0)M} = \xi^M$, which implies that

$$\delta^{(0)}_{\xi}\phi^M = -\Xi^M|_{\Theta=0} = -\xi^M. \tag{4.33}$$

Referring to Eq. (3.11), the above equation can be written in components, as follows:

$$\begin{pmatrix} \delta^{(0)}\phi^\mu \\ \delta^{(0)}\zeta^\alpha \end{pmatrix} = \begin{pmatrix} -\xi^\mu \\ -\epsilon^\alpha \end{pmatrix} = -\Xi^{(0)M}|_{\Theta=0} = -\xi^{(0)M}. \tag{4.34}$$

This equation shows that the leading-order transformation of the Stückelberg fields is a shift by the parameters $-\xi^{(0)M}$, which is characteristic of how Goldstone fields transform at the linear order. At linear order, $k = 1$, we have

$$\begin{aligned}\delta^{(1)}_{\xi}\phi^M \partial_M &= B_1^- \mathrm{ad}_{\delta_\phi} \delta_{\xi}|_{\Theta=0} = B_1^- [\delta_\phi, \delta_{\xi}]|_{\Theta=0} \\ &= B_1^- \delta_{\xi^{(1)}(\phi,\xi)}|_{\Theta=0} = -B_1^- \Xi^{(1)M}\partial_M|_{\Theta=0} \\ &= -B_1^- O^M{}_{\bar{N}}\xi^{(1)\bar{N}}\partial_M|_{\Theta=0} = -B_1^-\xi^{(1)M}\partial_M. \end{aligned} \tag{4.35}$$

We compute the commutator $[\delta_\phi, \delta_{\xi}]$ by looking at Eq. (3.25) and setting $\Xi^M_1$ to $\Phi^M$ and $\Xi^N_2$ to $\Xi^{N(0)} = \Xi^N$ in it. We get

$$\delta^{(1)}_{\xi}\phi^M = \frac{1}{2}(\Phi^N \partial_N \Xi^M - \Xi^N \partial_N \Phi^M)|_{\Theta=0}. \tag{4.36}$$

Note that the terms corresponding to the last two terms in Eq. (3.25) will vanish here at $\Theta = 0$, due to which they do not contribute here. The above expression can be written explicitly by substituting the expressions for $\Xi$ and $\Phi$ given in (3.6) and (4.2), respectively. Omitting the partial derivatives, we get

---

[9]$\delta_{\xi^{(k)}(\phi,\xi)}$ is not to be confused with $\delta^{(k)}_{\xi}\phi^M$. $\delta^{(k)}_{\xi}\phi^M \partial_M = -\frac{B_k^-}{k!}\delta_{\xi^{(k)}(\phi,\xi)}|_{\Theta=0}$.

$$\begin{pmatrix} \boldsymbol{\delta}^{(1)}\phi^\nu \\ \boldsymbol{\delta}^{(1)}\zeta^\alpha \end{pmatrix} = \begin{pmatrix} \frac{1}{2}(\phi^\mu \partial_\mu \xi^\nu - \xi^\mu \partial_\mu \phi^\nu) + i(\zeta\sigma^\mu\bar{\epsilon} - \epsilon\sigma^\mu\bar{\zeta}) \\ \frac{1}{2}(\phi^\mu \partial_\mu \epsilon^\alpha - \xi^\mu \partial_\mu \zeta^\alpha) + \frac{i}{2}(\epsilon\sigma^\mu\bar{\zeta} - \zeta\sigma^\mu\bar{\epsilon})\psi_\mu^\alpha \end{pmatrix}$$

$$= \frac{1}{2}\Xi^{(1)M}|_{\Theta=0} = \frac{1}{2}\boldsymbol{\xi}^{(1)M}. \quad (4.37)$$

This agrees with the results in [29].[10] At the next order in the Stückelberg fields, i.e., at $k=2$, we have

$$\delta_{\boldsymbol{\xi}}^{(2)}\boldsymbol{\phi}^M \partial_M = \frac{B_2^-}{2}\mathrm{ad}_{\delta_\phi}^2 \delta_\xi \Big|_{\Theta=0} = \frac{B_2^-}{2}[\delta_\phi, \delta_{\boldsymbol{\xi}^{(1)}(\phi,\xi)}]\Big|_{\Theta=0}$$
$$= \frac{B_2^-}{2}\delta_{\boldsymbol{\xi}^{(2)}(\phi,\xi)}\Big|_{\Theta=0} = -\frac{B_2^-}{2}\Xi^{(2)M}\partial_M\Big|_{\Theta=0}$$
$$= -\frac{B_2^-}{2}O^M{}_{\bar{N}}\boldsymbol{\xi}^{(2)\bar{N}}\partial_M\Big|_{\Theta=0} = -\frac{B_2^-}{2}\boldsymbol{\xi}^{(2)M}\partial_M. \quad (4.38)$$

Now we compute the commutator appearing above by setting $\Xi_1^M$ in Eq. (3.25) to $\Phi^M$ and $\Xi_2^N$ to $\Xi^{(1)N}$. It gives us

$$\delta_{\boldsymbol{\xi}}^{(2)}\boldsymbol{\phi}^M = \frac{1}{12}(\Phi^N \partial_N \Xi^{(1)M} - \Xi^{(1)N}\partial_N \Phi^M)\Big|_{\Theta=0}$$
$$+ \frac{1}{12}\frac{\partial \Xi^{(1)M}}{\partial \varphi_{sg}}\Big|_{\Theta=0}\delta_\phi \varphi_{sg}. \quad (4.39)$$

The last term above is nonzero since $\Xi^{(1)M}|_{\Theta=0} = \boldsymbol{\xi}^{(1)M}$ depends on the supergravity fields, as can be seen from Eq. (4.37). However, since $\Phi^M|_{\Theta=0}$ is independent of $\varphi_{sg}$ [cf. Eq. (4.3)], $(\partial\Phi^M/\partial\varphi_{sg})|_{\Theta=0}$ is 0. Then, $(\partial\Phi^M/\partial\varphi_{sg})|_{\Theta=0}$ corresponds to the last term in (3.25) and does not contribute here. Similarly, it does not contribute to $\delta_{\boldsymbol{\xi}}^{(k)}\boldsymbol{\phi}^M$ for any $k$.

Thus, we can set up a recursion procedure that allows us to compute the supergravity transformation of the Stückelberg fields at any order $k$ in $\boldsymbol{\phi}^{\bar{M}}$ and linearized in $\boldsymbol{\xi}^{\bar{M}}$. Recall that

$$\delta_{\boldsymbol{\xi}}^{(k)}\boldsymbol{\phi}^M = -\frac{B_k^-}{k!}\Xi^{(k)M}|_{\Theta=0}, \quad (4.40)$$

with

$$\Xi^{(k)M} = \Xi^{(k-1)N}\partial_N \Phi^M - \Phi^N \partial_N \Xi^{(k-1)M} - \frac{\partial \Xi^{(k-1)M}}{\partial \varphi_{sg}}\delta_\phi \varphi_{sg}. \quad (4.41)$$

Note that $B_k^- = 0$ for all odd $k$ except $k=1$. So,

$$\delta_{\boldsymbol{\xi}}\boldsymbol{\phi}^M = \delta_{\boldsymbol{\xi}}^{(0)}\boldsymbol{\phi}^M + \delta_{\boldsymbol{\xi}}^{(1)}\boldsymbol{\phi}^M + \cdots + \delta_{\boldsymbol{\xi}}^{(k)}\boldsymbol{\phi}^M + \ldots, \quad (4.42)$$

where $k$ is always even when $k > 1$.

Thus, we have shown the complete $\mathcal{N}=1$ supergravity action that allows for de Sitter solutions, obtained via the Stückelberg procedure [Eq. (4.6)], along with the supergravity transformation rules of all the fields appearing in the action [cf. (3.4), (3.21), Appendix A, and (4.30)].

## V. DISCUSSION

In this work, we continued our program of constructing the superspace version of unimodular gravity via a super-Stückelberg mechanism, which we started in [29,38]. In this construction, the unimodularity condition requires that the chiral density superfield is constrained via a chiral superfield Lagrange multiplier. General coordinate invariance and local supersymmetry are then restored using the super-Stückelberg trick that we introduced perturbatively in [29]. In the present work, we demonstrated the process of restoring general coordinate invariance and local supersymmetry through the super-Stückelberg mechanism to all orders in the Stückelberg fields. To do this, we first introduced finite supergravity transformation in Sec. III A. We used this tool to find the expressions for the supergravity transformation of the Stückelberg fields, given by Eq. (4.30). With this result in hand, we found recursive expressions for the transformation of the Stückelberg fields to all orders in $\boldsymbol{\phi}^{\bar{M}}$, given by Eq. (4.40). We used this result to compute the zeroth and first-order transformations, to show that they are consistent with our previous perturbative result in [29], as expected.

This work opens up several new directions in which we can extend the ideas. The first is, once again, inspired by standard unimodular gravity, in which the Stückelberg formulation is more elegantly presented in terms of a three-form field and the corresponding four-form field strength, as originally proposed by Heneaux and Teitelboim [49]. The extension to supergravity is nontrivial, not least because three-form multiplets are usually introduced by placing the corresponding field strength in the F-term of a chiral superfield [61].

With our present construction to all orders now at hand, it would be interesting to compare with the brane and constrained superfield constructions of Bandos *et al.* [20] and those of [14–18], respectively. We have already shown that the two formulations agree up to cubic order in the Stückelberg fields in our previous work [38]. If they agree to all orders, it could point to a more natural construction at higher energies, or with more supersymmetry. If they disagree, it would be interesting to explore the phenomenological differences, particularly with a view to describing the late time dynamics of our Universe. Such comparisons could provide insights into the formulation's robustness and potential unification implications for theories incorporating supersymmetry at higher scales.

---

[10]The additional terms in (4.37), relative to [29], arise from the fact that we are considering the complete Lagrangian here, whereas in [29] we truncated to the second order in the fermions.

Finally, we note that one of the original motivations for this program, at least for two of us, was the desire to build the model of vacuum energy sequestering [39–45] into a supersymmetric framework. VES is a low-energy effective theory describing general relativity with global constraints. These global constraints ensure that radiative corrections to the vacuum energy are kept under control, presenting a possible solution to the cosmological constant problem. The theory is an extension of Henneaux-Teitelboim's formulation of unimodular gravity, where the Planck mass also behaves as a Lagrange multiplier, directly coupling to a second species of three-form. To extend this to a theory of supergravity, perhaps it is enough to promote the Planck mass to a chiral superfield like we have for the cosmological constant, coupling it to another set of Stückelberg fields in the same way. A supersymmetric theory of VES could be a route toward building the model into a more fundamental theory. The resulting theory may yield interesting phenomenological consequences, offering new tests of the VES mechanism.

## ACKNOWLEDGMENTS

We are grateful to Javier Peraza and Giorgio Pizzolo for useful discussions. S. B. is supported by the Austrian Science Fund, FWF, Project No. P34562. S. N. is partially supported by an STFC Consolidated Grant, No. ST/T000708/1. A. P. is partially supported by the STFC Consolidated Grants No. ST/V005596/1, No. ST/T000732/1, and No. ST/X000672/1. I. Z. is partially funded by STFC Grant No. ST/X000648/1.

## DATA AVAILABILITY

No data were created or analyzed in this study.

## APPENDIX A: INFINITESIMAL SUPERGRAVITY TRANSFORMATION OF SUPERGRAVITY FIELDS

The fields belonging to the minimal supergravity multiplet are $\varphi_{sg} = (e_\mu^a, \psi_\mu^\alpha, b_\mu, M)$. They transform under infinitesimal diffeomorphism + local supersymmetry transformation as follows:

$$\varphi_{sg} \to \varphi_{sg} + \delta_\xi \varphi_{sg} = \varphi_{sg} + \delta_\xi \varphi_{sg} + \delta_{(\epsilon,\bar\epsilon)} \varphi_{sg}, \quad (A1)$$

where

$$\begin{aligned}
\delta_\xi e_\mu^a &= -\xi^\nu \partial_\nu e_\mu^a - (\partial_\mu \xi^\nu) e_\nu^a, \\
\delta_\xi \psi_\mu^\alpha &= -\xi^\nu \partial_\nu \psi_\mu^\alpha - (\partial_\mu \xi^\nu) \psi_\nu^\alpha, \\
\delta_\xi b_\mu &= -\xi^\nu \partial_\nu b_\mu - (\partial_\mu \xi^\nu) b_\mu, \\
\delta_\xi M &= -\xi^\mu \partial_\mu M,
\end{aligned} \quad (A2)$$

and

$$\begin{aligned}
\delta_{(\epsilon,\bar\epsilon)} e_\mu^a &= i(\psi_\mu \sigma^a \bar\epsilon - \epsilon \sigma^a \bar\psi_\mu), \\
\delta_{(\epsilon,\bar\epsilon)} e_a^\mu &= i(\epsilon \sigma^\mu \bar\psi_a - \psi_a \sigma^\mu \bar\epsilon), \\
\delta_{(\epsilon,\bar\epsilon)} \psi_\mu^\alpha &= -2\mathcal{D}_\mu \epsilon^\alpha + \frac{i}{3} M (\epsilon \sigma_\mu \bar\epsilon)^\alpha + i b_\mu \epsilon^\alpha + \frac{i}{3} b^\nu (\epsilon \sigma_\nu \bar\sigma_\mu)^\alpha, \\
\delta_{(\epsilon,\bar\epsilon)} \bar\psi_{\mu\dot\alpha} &= -2\mathcal{D}_\mu \bar\epsilon_{\dot\alpha} - \frac{i}{3} M^* (\epsilon \sigma_\mu)_{\dot\alpha} - i b_\mu \bar\epsilon_{\dot\alpha} - \frac{i}{3} b^\nu \bar\epsilon_{\dot\beta} (\bar\sigma_\nu \sigma_\mu)^{\dot\beta}{}_{\dot\alpha}, \\
\delta_{(\epsilon,\bar\epsilon)} M &= -\epsilon(\sigma^\mu \bar\sigma^\nu \psi_{\mu\nu} + i b^\mu \psi_\mu - i \sigma^\mu \bar\psi_\mu M), \\
\delta_{(\epsilon,\bar\epsilon)} M^* &= \bar\epsilon(-\bar\sigma^\mu \sigma^\nu \bar\psi_{\mu\nu} + i b^\mu \bar\psi_\mu + i \bar\sigma^\mu \psi_\mu M^*), \\
\delta_{(\epsilon,\bar\epsilon)} b_{\alpha\dot\alpha} &= \epsilon \mathcal{F}(\psi; M, b) \\
&= \epsilon^\delta \left[ \frac{3}{4} \bar\psi_\alpha{}^{\dot\gamma}{}_{\delta\dot\gamma\dot\alpha} + \frac{1}{4} \varepsilon_{\delta\alpha} \bar\psi^{\gamma\dot\gamma}{}_{\gamma\dot\alpha\dot\gamma} - \frac{i}{2} M^* \psi_{\alpha\dot\alpha\delta} + \frac{i}{4} (\bar\psi_{\alpha\dot\rho}{}^{\dot\rho} b_{\delta\dot\alpha} + \bar\psi_{\delta\dot\rho}{}^{\dot\rho} b_{\alpha\dot\alpha} - \bar\psi_{\delta}{}^{\dot\rho}{}_{\dot\alpha} b_{\alpha\dot\rho}) \right] \\
&\quad - \bar\epsilon^{\dot\delta} \left[ \frac{3}{4} \psi^\gamma{}_{\dot\delta\dot\gamma\alpha\dot\alpha} + \frac{1}{4} \varepsilon_{\dot\delta\dot\alpha} \psi_\alpha{}^{\dot\gamma\gamma}{}_{\dot\gamma\gamma} + \frac{i}{2} M \bar\psi_{\alpha\dot\alpha\dot\delta} - \frac{i}{4} (\psi_{\rho\dot\alpha}{}^\rho b_{\alpha\dot\delta} + \psi_{\rho\dot\delta}{}^\rho b_{\alpha\dot\alpha} - \psi^\rho{}_{\dot\delta\alpha} b_{\rho\dot\alpha}) \right].
\end{aligned} \quad (A3)$$

## APPENDIX B: USEFUL IDENTITIES WITH PROOFS

In this appendix, we collect a series of identities, and their proofs, which we use throughout the main text. These identities can be found in [56], while some of the proofs are expanded or added here. Throughout this appendix, $Z^M = (x^\mu, \Theta^\alpha)$ is the chiral superspace coordinate.

*Identity B1*. For an arbitrary superfunction $\tau(Z)$, the operator

$$[K, \tau(Z)] = K(\tau(Z)), \tag{B1}$$

where $K = K^M(Z)\partial_M = K^m(x,\Theta)\partial_m + K^\mu(x,\Theta)\partial_\mu$.

*Proof.* We take a test function $\chi(Z)$ of supercoordinate $Z^M$. We have

$$\begin{aligned}
K(\tau(Z)\chi(Z)) &= K^M \partial_M(\tau(Z)\chi(Z)) \\
&= (K^M \partial_M \tau)\chi + (-1)^{\varepsilon_\tau \varepsilon_M} \tau(K^M \partial_M \chi) \quad \text{(by Leibniz rule)} \\
\Rightarrow K^M \partial_M(\tau\chi) - \tau(K^M \partial_M \chi) &= (K^M \partial_M \tau)\chi \\
\Rightarrow K(\tau\chi) - \tau(K\chi) &= (K\tau)\chi \\
\Rightarrow [K, \tau(Z)]\cdot &= K(\tau(Z))\cdot \quad [\because \chi(Z) \text{ can be any arbitrary function}].
\end{aligned}$$

∎

*Identity B2*. For an arbitrary superfunction $\tau(Z)$,

$$e^{-K}\tau(Z)e^K = e^{-K}(\tau(Z)). \tag{B2}$$

*Proof.* From Proposition 3.35 in [53], we know that

$$e^{-K}\tau(Z)e^K = e^{-\text{ad}_K}\tau(Z),$$

where $\text{ad}_K \tau(Z) = [K, \tau(Z)]$. Plugging the alternate expression for $[K, \tau(Z)]$ from identity (B1) into the above equation, we prove identity (B2). ∎

*Identity B3*. For a superanalytic function $\rho(Z)$,

$$\rho(e^{-K}Ze^K) = e^{-K}\rho(Z)e^K. \tag{B3}$$

*Proof.* Since a superanalytic function can always be expressed as a Taylor series around any point $Z_0^M$ in its domain, $\rho(Z)$ being a superanalytic function can be expressed as the following Taylor series expanded around 0:

$$\rho(Z) = \sum_{k=0}^{k=\infty} a_{M_1 M_2 \ldots M_k} Z^{M_1} Z^{M_2} \ldots Z^{M_k}.$$

Then,

$$\begin{aligned}
\rho(e^{-K}Ze^K) &= \sum_{k=0}^{k=\infty} a_{M_1 M_2 \ldots M_k} e^{-K} Z^{M_1} e^K e^{-K} Z^{M_2} e^K \ldots e^{-K} Z^{M_k} e^K \\
&= \sum_{k=0}^{k=\infty} a_{M_1 M_2 \ldots M_k} e^{-K} Z^{M_1} Z^{M_2} \ldots Z^{M_k} e^K \\
&= e^{-K} \rho(Z) e^K.
\end{aligned}$$

∎

*Identity B4*. For a superanalytic function, $\rho(Z)$,

$$\rho(e^{-K}Ze^K) = e^{-K}(\rho(Z)). \tag{B4}$$

*Proof.* Substituting identity (B3) into identity (B2), we get identity (B4). ∎

*Identity B5*. Under the finite passive transformation $Z \to Z' = e^{-K}Z$, a superanalytic function $\rho(Z)$ transforms as

$$\rho'(Z') = e^{-K}\rho'(Z). \tag{B5}$$

*Proof.* Under $Z \to Z' = e^{-K}Z$,

$$\begin{aligned}
\rho'(Z') &= \rho'(e^{-K}Z) \\
&= \rho'(e^{-K}Ze^K) \\
&= e^{-K}\rho'(Z),
\end{aligned}$$

where in the second line we took $\rho(Z) = Z$ in (B4). ∎

*Identity B6*. The finite passive transformation $Z \to Z' = e^{-K}Z$ induces the following finite active transformation on a superanalytic function $\rho(Z)$:

$$\rho'(Z) = e^K \rho(Z). \tag{B6}$$

*Proof.* We know that under the finite passive transformation $Z \to Z' = e^{-K}Z$, a superanalytic function $\rho(Z)$ obeys [cf. Eq. (1.11.6) in [56]]

$$\begin{aligned}
\rho(Z) &= \rho'(Z') \\
&= e^{-K}\rho'(Z),
\end{aligned}$$

where in the second line we used (B5). The above result gives identity (B6).[11] ∎

---

[11]Also proved in Eq. (1.11.8) in [56].

*Identity B7.* When $Z$ transforms as $Z \to Z' = e^{-K}Z$,

$$\mathrm{Ber}\left(\frac{\partial Z'^M}{\partial Z^N}\right) = 1 \cdot e^{-\tilde{K}}, \tag{B7}$$

where $\tilde{K} = K^M \overleftarrow{\partial}_M = (-1)^M \overleftarrow{\partial}_M K^M + (-1)^M \partial_M K^M$.

*Proof.* See Sec. 1.11.3 in [56]. ∎

*Identity B8.* For arbitrary superfunctions $\tau(Z)$ and $\chi(Z)$,

$$\int d^6 Z \, \tau(Z) \cdot e^{-\tilde{K}} \chi(Z) = \int d^6 Z \, \tau(Z) e^K \chi(Z). \tag{B8}$$

*Proof.* We know that

$$\int d^6 Z \, \tau(Z) \cdot e^{-\tilde{K}} \chi(Z) = \int d^6 Z \, \tau(Z) \cdot \frac{(-\tilde{K})^n}{n!} \chi(Z). \tag{B9}$$

We evaluate every term on the rhs of the above equation one-by-one. The first term is simplified as $\int d^6 Z \, \tau \chi$. The second term is

$$-\int d^6 Z \, \tau(Z) \cdot \tilde{K} \chi(Z)$$
$$= -\int d^6 Z \, \tau \cdot K^M \overleftarrow{\partial}_M \chi$$
$$= -\int d^6 Z \, (\tau K^M) \cdot \overleftarrow{\partial}_M \chi$$
$$= -\int d^6 Z \, (-1)^M \partial_M (\tau K^M) \chi$$
$$= -(-1)^M \int d^6 Z \, \partial_M (\tau K^M \chi) - (-1)^M \tau K^M \partial_M \chi$$
$$= \int d^6 Z \, \tau K \chi.$$

In the penultimate line, the total derivative term vanishes after integration. The third term on the rhs of Eq. (B9) is

$$\int d^6 Z \, \tau(Z) \cdot \frac{1}{2!} \tilde{K}^2 \chi(Z) = \frac{1}{2} \int d^6 Z \, (\tau \cdot K^N \overleftarrow{\partial}_N) \cdot K^M \overleftarrow{\partial}_M \chi$$
$$= \frac{1}{2} \int d^6 Z \, \{(-1)^N \partial_N (\tau K^N)\} \cdot K^M \overleftarrow{\partial}_M \chi$$
$$= \frac{(-1)^N}{2} \int d^6 Z \, \{\partial_N (\tau K^N) K^M\} \cdot \overleftarrow{\partial}_M \chi$$
$$= \frac{(-1)^{(N+M)}}{2} \int d^6 Z \, \partial_M \{\partial_N (\tau K^N) K^M\} \chi$$
$$= \frac{(-1)^{(N+M)}}{2} \int d^6 Z \, \partial_M \{\partial_N (\tau K^N) K^M \chi\} - (-1)^M \partial_N (\tau K^N) K^M \partial_M \chi$$
$$= \frac{(-1)^{(N+M)}}{2} \int d^6 Z \, - (-1)^M \partial_N \{(\tau K^N) K^M \partial_M \chi\} + (-1)^{(M+N)} \tau K^N \partial_N K^M \partial_M \chi$$
$$= \frac{1}{2} \int d^6 Z \, \tau K^2 \chi.$$

Thus, evaluating the $n$th-order term on the rhs of Eq. (B9), we get

$$\int d^6 Z \, \tau(Z) \cdot \frac{(-\tilde{K})^n}{n!} \chi(Z) = \int d^6 Z \, \tau \frac{K^n}{n!} \chi.$$

On summing up all these terms, we prove identity (B8). ∎

## APPENDIX C: ALTERNATIVE DERIVATION OF THE SUPERGRAVITY TRANSFORMATION OF STÜCKELBERG FIELDS

In this appendix, we give an alternative derivation for the results in Sec. IV, and in particular Eq. (4.28), describing the transformation rule for our Stückelberg fields. This provides a sanity check for our results. We start by generalizing the following result from unimodular gravity: the Stückelberg-dressed coordinate (i.e., the object obtained by performing a coordinate transformation, and then promoting the parameters to fields) should transform like a set of $d$ scalars, with $d$ being the dimension of our spacetime. In our case, for $d = 4$ and $\mathcal{N} = 1$ supergravity in the chiral formulation, the Stückelberg-dressed supercoordinate should transform like a set of $4 + 2$ chiral superfields.

The finite transformation of the supercoordinates is given in (4.9). After performing the Stückelberg replacement (4.1), we denote

$$Y^M \equiv e^{-\delta_\phi} Z^M. \tag{C1}$$

At the linear level in $\xi^{\bar{M}}$, the above transforms, under an active transformation, as

$$\delta Y^M = \delta(e^{-\delta_\phi} Z^M) = -\mathcal{O}_{\delta_\phi}(\delta'_\phi - \delta_\phi) e^{-\delta_\phi} Z^M$$
$$= -\mathcal{O}_{\delta_\phi}(\delta'_\phi - \delta_\phi) Y^M, \tag{C2}$$

where we used the variation of the exponential map

$$\delta(e^{-X}) = -\mathcal{O}_X(\delta X) e^{-X}, \tag{C3}$$

with

$$\mathcal{O}_X = \frac{1 - e^{-\mathrm{ad}_X}}{\mathrm{ad}_X} = \sum_{k=0}^\infty \frac{(-1)^k}{(k+1)!} \mathrm{ad}_X^k. \tag{C4}$$

Using (4.25), we can rearrange (C2) as

$$\delta Y^M = \mathcal{O}_{\delta_\phi} O^P{}_{\bar{N}} \delta_\xi \phi^{\bar{N}} \partial_P Y^M + \mathcal{O}_{\delta_\phi} \mathrm{ad}_{\delta_\phi} \delta_\xi Y^M$$
$$= \mathcal{O}_{\delta_\phi} O^P{}_{\bar{N}} \delta_\xi \phi^{\bar{N}} \partial_P Y^M + \frac{1 - e^{-\mathrm{ad}_{\delta_\phi}}}{\mathrm{ad}_{\delta_\phi}} \mathrm{ad}_{\delta_\phi} \delta_\xi Y^M$$
$$= \mathcal{O}_{\delta_\phi} O^P{}_{\bar{N}} \delta_\xi \phi^{\bar{N}} \partial_P Y^M + \delta_\xi Y^M - e^{-\mathrm{ad}_{\delta_\phi}} \delta_\xi Y^M. \tag{C5}$$

On the other hand, as we said, we need $Y^M$ to transform like a chiral superfield, which, in our notation, is just

$$\delta Y^M = \delta_\xi Y^M. \tag{C6}$$

Comparing (C5) with (C6), we get

$$\mathcal{O}_{\delta_\phi} O^P{}_{\bar{N}} \delta_\xi \phi^{\bar{N}} \partial_P Y^M = e^{-\mathrm{ad}_{\delta_\phi}} \delta_\xi Y^M. \tag{C7}$$

We can invert $\mathcal{O}_{\delta_\phi}$ to write

$$O^P{}_{\bar{N}} \delta_\xi \phi^{\bar{N}} \partial_P Y^M = \frac{\mathrm{ad}_{\delta_\phi}}{1 - e^{-\mathrm{ad}_{\delta_\phi}}} e^{-\mathrm{ad}_{\delta_\phi}} \delta_\xi Y^M. \tag{C8}$$

The above is true for any $Y^M$, so, after some rearrangement, we have

$$O^P{}_{\bar{N}} \delta_\xi \phi^{\bar{N}} \partial_P = \frac{\mathrm{ad}_{\delta_\phi}}{e^{\mathrm{ad}_{\delta_\phi}} - 1} \delta_\xi, \tag{C9}$$

which matches with (4.28).